\documentclass[lettersize,journal]{IEEEtran}
\usepackage{amsmath,amsfonts}
\usepackage{algorithmic}
\usepackage{algorithm}
\usepackage{array}
\usepackage[caption=false,font=normalsize,labelfont=sf,textfont=sf]{subfig}
\usepackage{textcomp}
\usepackage{stfloats}
\usepackage{url}
\usepackage{verbatim}
\usepackage{graphicx}
\usepackage{cite}
\usepackage{makecell}
\usepackage{pifont}
\usepackage{multirow}
\usepackage{multicol}

\begin{document}

%\title{Scheduling Communication Resources in \\ Wireless IoT Networks to Maximize the Energy Efficiency of Federated Learning}
\title{Improving Energy Efficiency in Federated Learning Through the Optimization of Communication Resources Scheduling of Wireless IoT Networks}

\author{Renan R. de Oliveira, Kleber V. Cardoso, and Antonio Oliveira-Jr
        % <-this % stops a space
%\thanks{This paper was produced by the IEEE Publication Technology Group. They are in Piscataway, NJ.}% <-this % stops a space
\thanks{Renan R. de Oliveira, Federal University of Goiás, Goiânia-GO and Federal Institute of Goiás, Goiânia-GO, e-mail: renan.rodrigues@ifg.edu.br. 
Kleber V. Cardoso, Federal University of Goiás, Goiânia-GO, e-mail: kleber@ufg.br.
Antonio Oliveira-Jr, Federal University of Goiás, Goiânia-GO and Fraunhofer Portugal AICOS, Porto-Portugal, e-mail: antoniojr@ufg.br.}}

% The paper headers
%\markboth{Journal of \LaTeX\ Class Files,~Vol.~14, No.~8, August~2021}%
%{Shell \MakeLowercase{\textit{et al.}}: A Sample Article Using IEEEtran.cls for IEEE Journals}

%\IEEEpubid{0000--0000/00\$00.00~\copyright~2021 IEEE}
% Remember, if you use this you must call \IEEEpubidadjcol in the second
% column for its text to clear the IEEEpubid mark.

\maketitle

\begin{abstract}
Federated Learning (FL) allows devices to train a global machine learning model without sharing data. In the context of wireless networks, the inherently unreliable nature of the transmission channel introduces delays and errors that compromise the regularity of updating the global model. Furthermore, limited resources and energy consumption of devices are factors that affect FL performance. Therefore, this work proposes a new FL algorithm called FL-E2WS that considers both the requirements of federated training and a wireless network within the scope of the Internet of Things. To reduce the energy cost of devices, FL-E2WS schedules communication resources to allocate the ideal bandwidth and power for the transmission of models under certain device selection and uplink resource block allocation, meeting delay requirements, power consumption, and packet error rate. The simulation results demonstrate that FL-E2WS reduces energy consumption by up to 70.12\% and enhances the accuracy of the global model by up to 10.21\% compared to the FL algorithms that lacks transmission channel knowledge. Additionally, when compared to FL versions that scale communication resources, FL-E2WS achieves up to a 38.61\% reduction in energy consumption and improves the accuracy of the global model by up to 1.61\%.
\end{abstract}


\section{Introduction}
\IEEEPARstart{P}ROJECTIONS for B5G/6G networks indicate a scenario characterized by emerging intelligent applications, enabling Machine Learning (ML) models to be executed on heterogeneous and resource-limited edge devices \cite{Stergiou}. However, when it comes to training, one assumption is that these models must be trained centrally in the cloud using training data from many devices \cite{Beutel}. On the other hand, transmitting massive amounts of data to the cloud can compromise data privacy, increase traffic volume, and overload communication \mbox{networks \cite{Xiaopeng}}.

Due to the need to process a massive amount of private data continuously generated by devices such as smartphones, wearable devices, autonomous vehicles, and other Internet of Things devices (IoT), training ML models on edge devices has sparked increasing interest in recent years. In this context, FL \cite{McMahan} was introduced as a decentralized ML approach that allows devices to collaboratively train a shared model while keeping data private across devices. In this approach known as centralized FL \cite{Martinez}, only the parameters of the locally trained models are shared with the aggregator server.

FL over wireless networks has advantages over centralized ML because transmitting ML model parameters instead of training data between devices and the Base Station (BS) can save energy and network resources and reduce communication latency \cite{Zhaohui}. Furthermore, FL contributes to preserving data privacy as training data remains on devices. In this way, edge computing integrated into the network infrastructure, as proposed in the Multi-Access Edge Computing (MEC) approach, contributes to the deployment of FL parameter servers at the edge of the network, that is, closer to the devices where the models are trained \cite{Hammoud}. FL combined with MEC can facilitate the involvement of massive ubiquitous devices in mobile networks as an enabler for emerging\mbox{ applications \cite{Duan}}. Furthermore, efficient collaboration between aggregator servers at the edge and devices participating in FL is crucial to optimizing the performance of the global model \cite{Sun}.

However, FL approaches in wireless networks must consider the occurrence of failures in the distributed training process due to communication resource constraints and the inherent unreliability of the wireless channels \cite{Chen-Walid}. In this case, the presence of limited bandwidth, dynamic wireless channels and high interference can result in significant transmission latency, which can negatively impact real-time applications such as wireless virtual reality (VR), device to device (D2D) and vehicle to vehicle (V2V) communications \cite{Deng}. Furthermore, according to \cite{Zhu-Guangxu}, a typical strategy to achieve satisfactory performance in FL is to schedule as many devices as possible in each communication round. However, it is not desirable for all devices to send their local models to the parameter server, especially when updates are transmitted over a resource-limited wireless connection \cite{Chen-Hao}.

On the other hand, the energy consumption of the devices and the convergence speed of the global model are important factors that affect the performance of FL algorithms. According to \cite{Peng}, when only energy consumption is considered, the training process can take longer and the system efficiency can be harmed. Conversely, when only the convergence rate is considered, devices with limited communication resources tend to consume an excessive amount of energy to transmit models. In this context, the work of \cite{Wahab} and \cite{Soltani} highlights the importance of designing models to determine the ideal number of clients that need to participate in each training iteration.

To overcome these shortcomings, this work proposes a new FL algorithm called \mbox{FL-E2WS}\footnote{Available at https://github.com/LABORA-INF-UFG/FL-E2WS} (\textit{Federated Learning - Energy-efficient Wireless Scheduling}), characterized as an optimization problem, whose objective is to minimize the energy consumption of devices in a wireless IoT network without compromising the quality of the global model. \mbox{FL-E2WS} addresses the problem into two subproblems. The former is called the device selection problem, which aims to select (from a subset of devices) those participants who will maximize the amount of data used during the training of local models for the next communication round. The latter, called the communication resource scheduling problem, is treated as a designation problem that prioritizes the minimization of energy consumption and considers the ideal bandwidth and power for transmitting the models to the BS under a given device selection and allocation of \textit{uplink} RBs, meeting delay, energy consumption, and packet error rate requirements. The central contributions of this article are listed below:

\begin{itemize} 
\item \textbf{FL formulation in wireless IoT networks:} FL-E2WS uses formulations that define wireless IoT network \mbox{simulation} models for FL tasks by introducing attributes (such as power and bandwidth) used to optimize energy efficiency.

\item \textbf{New approach:} FL-E2WS improves the representativeness of locally trained models by selecting devices (from a subset of devices) in each communication round based on maximizing the amount of data from the devices. Then, FL-E2WS scales communication resources, minimizing the energy consumption of model transmission while maximizing the number of devices selected for each round of communication.

\item \textbf{Efficient exact solution}: FL-E2WS scales communication resources based on the Mixed Integer Linear Programming (MILP) technique using a mathematical optimization modeling library (i.e. PuLP), despite the complexity of the problem.

\item \textbf{Comparison with other FL algorithms: } FL-E2WS is compared with FedAvg \cite{McMahan} and POC \cite{POC}. Additionally, with the aim of ensuring fairness in the comparison with \mbox{FL-E2WS}, variations of FedAvg and POC were implemented, referred to in this work as FedAvg-w$_{Opt}$ and POC-w$_{Opt}$, incorporating the knowledge of the wireless network using the MILP algorithm of \mbox{FL-E2WS} for the assignment of uplink  resource blocks (RBs) and operating with the fixed bandwidth and transmission power allocation.

\item \textbf{Evaluation and new insights:} Experimental results are presented that demonstrate that the \mbox{FL-E2WS} algorithm outperforms its equivalents by achieving an accuracy with a slightly higher tendency and a significant reduction in the energy consumption of the devices without compromising the quality of the global model. 

\end{itemize}

%In addition to this introductory section, the remainder of this article is organized into sections, as described below. 

The rest of the paper is organized as follows. Section \ref{related-work} discusses related work. Section \ref{model} presents the mathematical models that were used in simulating the wireless IoT network for FL tasks. Section \ref{problem-formulation} describes the formulation of \mbox{FL-E2WS}. Section \ref{results} discusses the simulation results and analysis. Finally, Section \ref{conclusions} presents the final considerations and indicates directions for future work.

\section{Related Work}
\label{related-work}

Since the original proposition of FL, an increasing number of studies dedicated to improving this paradigm have emerged. However, many of these works have focused on minimizing computing costs or optimizing parameter update strategies and did not address the problems related to implementing FL while considering the uncertainties inherent to wireless channels and minimizing the energy consumption of devices.

In the initial formulation of FL proposed by \cite{McMahan}, the authors introduced the concept of FL and presented the performance of the Federated Averaging (\mbox{FedAvg}) algorithm with different configurations, datasets, and ML models. However, the work did not provide a comprehensive discussion in the context of wireless networks. In \cite{POC}, the authors present the Power-Of-Choice (POC) algorithm, which selects devices for the next round of communication based on the largest values of the local loss function. As an alternative, the authors suggest that the server uses accumulated average loss information previously sent by devices along with local models. In \cite{Asad}, different FL strategies are evaluated in terms of communication accuracy and efficiency. The authors proposed a strategy based on data sharing to address the distribution of non-IID data. However, the works of \cite{POC} and \cite{Asad} focused on designing algorithms to improve learning convergence, without addressing other factors that affect FL performance, such as wireless communication and minimizing energy consumption due to the limited nature of the devices.

In \cite{Cao} the authors present an overview of current methodologies for distributed learning in terms of communication. The authors point out that FL presents significant challenges due to communication constraints and the dynamic nature of wireless networks. Furthermore, the importance of device selection and optimization of communication resource allocation to improve FL efficiency in wireless networks is highlighted. In \cite{Tran}, the authors developed one of the first works that considered both FL metrics and factors inherent to wireless networks. The article presents an FL model in wireless networks considering computing and communication characteristics. Then, a minimization problem for FL tasks is formulated with the objective of optimizing the training time of the devices local models while taking into account the minimization of energy consumption. However, the proposed model considers the participation of a random sampling of device subsets in each communication round and does not evaluate the effects of implementing a device selection policy.

The work by \cite{Chen-Walid} presents an FL model in wireless networks and a proposal for device selection and resource allocation jointly through the formulation of an optimization problem that aims to minimize the training loss of the global model. Similarly, in \cite{Chen-Hao} the authors study investigated communication resource optimization for FL, where the problem was divided into the subproblems of device scheduling and resource allocation. The scaling policy considered the reuse of obsolete parameters from local models to reduce communication costs. However, the research by \cite{Chen-Walid} and \cite{Chen-Hao} does not comprehensively address scalability for wireless IoT network environments, as the results were obtained based on iterations of fewer than a few devices and exclusively used the standard version of MNIST as a reference dataset.

The problem of communication efficiency and minimizing transmission time has been addressed in several FL. In the article by \cite{Beitollahi}, a method for scheduling focused on energy efficiency of FL aggregator servers with energy constraints is proposed. The problem is formulated as mixed-integer non-linear problem (MINLP) considering resource allocation and user selection with a focus on transmission time efficiency of local models. In \cite{Perazzone} a device selection and power allocation algorithm is presented that minimizes the average communication time required to achieve a target accuracy in federated learning. The approach uses stochastic optimization to determine the optimal selection probabilities and transmission powers, considering the varying conditions of the communication channel. These works focus on communication efficiency and minimizing transmission time, which contributes to reducing energy consumption. However, these works do not use a strategy for selecting a device subset for the stage of optimizing devices communication resources.

According to \cite{Shi}, the same optimization problem can have different characteristics from different perspectives, and the same application can be formulated as different optimization problems. In this context, tables \ref{tab:related-work} and \ref{tab:related-work-2} present a summary of related works and their respective target of optimization.

\begin{table*}[h]
\caption{Related Work\label{tab:related-work}}
\centering
\begin{tabular}{c || c c c c c c c}
\hline
\multirow{2}{*}{\textbf{Ref.}} & \multirow{2}{*}{\makecell{\textbf{Wireless} \\ \textbf{Networks}}} & \multicolumn{2}{c}{\textbf{Device Selection Scheme}} & \multirow{2}{*}{\makecell{\textbf{Number of Devices} \\ ($\geq 100$)}}  & \multirow{2}{*}{\makecell{\textbf{Prioritize} \\ \textbf{Energy Efficiency}}} & \multirow{2}{*}{\makecell{\textbf{Specific} \\ \textbf{Optimization Method}}}
 \\ 
 &  & \textbf{Initial Non-Random} & \makecell{\textbf{With Optimization
}} & &  
 \\ 
\hline
\cite{McMahan}  & \ding{55} & \ding{55} & \ding{55} & \scalebox{1.2}{\checkmark} & \ding{55} & \ding{55} \\
\hline
\cite{POC}  & \ding{55} & \scalebox{1.2}{\checkmark}  & \ding{55} & \scalebox{1.2}{\checkmark} & \ding{55} & \ding{55} \\
\hline
\cite{Asad}  & \ding{55} & \scalebox{1.2}{\checkmark}  & \ding{55} & \scalebox{1.2}{\checkmark} & \ding{55} & \ding{55} \\
\hline
\cite{Cao}  & \scalebox{1.2}{\checkmark} & \scalebox{1.2}{\checkmark}  & \scalebox{1.2}{\checkmark} & \ding{55} & \ding{55} & \ding{55} \\
\hline
\cite{Tran}  & \scalebox{1.2}{\checkmark} & \ding{55}  & \scalebox{1.2}{\checkmark} & \ding{55} & \scalebox{1.2}{\checkmark} & Pareto Model \\
\hline
\cite{Chen-Walid} & \scalebox{1.2}{\checkmark} & \ding{55}  & \scalebox{1.2}{\checkmark} & \ding{55} & \ding{55} & Hungarian Algorithm \\
\hline
\cite{Chen-Hao} & \scalebox{1.2}{\checkmark} & \ding{55}  & \scalebox{1.2}{\checkmark} & \ding{55} & \ding{55} & Lagrange Multiplier \\
\hline
\cite{Beitollahi}  & \scalebox{1.2}{\checkmark} & \ding{55} & \scalebox{1.2}{\checkmark} & \ding{55} & \scalebox{1.2}{\checkmark} & Stochastic Optimization \\
\hline
\cite{Perazzone}  & \scalebox{1.2}{\checkmark} & \ding{55} & \scalebox{1.2}{\checkmark} & \scalebox{1.2}{\checkmark} & \scalebox{1.2}{\checkmark} & MINLP \\
\hline
FL-E2WS (Ours)  & \scalebox{1.2}{\checkmark} & \scalebox{1.2}{\checkmark} & \scalebox{1.2}{\checkmark}  & \scalebox{1.2}{\checkmark} & \scalebox{1.2}{\checkmark} & MILP \\
\hline
\end{tabular}
\end{table*}

\begin{table*}[h]
\caption{Target of optimization of related work
\label{tab:related-work-2}}
\centering
\begin{tabular}{c||c}
 \hline
\textbf{Ref.} & \textbf{Target of Optimization} 
 \\
 \hline
\cite{Tran}  & \makecell{Resource allocation to optimize the convergence time and power consumption.}  \\
\hline
\cite{Chen-Walid} & \makecell{Adjust the optimal transmit power for each device under a block allocation scheme of uplink RBs to minimize the FL loss function.} \\ 
\hline
\cite{Chen-Hao} & \makecell{Optimization of the allocation of communication resources aiming at communication efficiency and FL model accuracy.} \\ 
\hline
\cite{Beitollahi} & \makecell{Resource scheduling focused on energy efficiency of FL aggregator servers with energy constraints.} \\ 
\hline
\cite{Perazzone} & \makecell{Power allocation aiming to minimize the average communication time required to achieve a target accuracy in FL.}  \\
\hline
FL-E2WS (Ours)  & \makecell{Improve device energy efficiency while maximizing the number of devices selected in each communication round.}  \\
\hline
\end{tabular}
\end{table*}

Motivated by the gaps identified in related work, \mbox{FL-E2WS} is proposed as a new FL algorithm that considers both factors inherent to the wireless environment and factors that influence the global model learning process, taking into account the minimization of energy consumption of devices in the model transmission stage without compromising the quality of the global model, while maximizing the number of devices selected for each communication round. The evaluation of FL-E2WS involves a slightly higher device density compared to the works cited in the context of wireless networks and uses variations of MNIST and Fashion-MNIST that make the data heterogeneous, reflecting the intrinsic characteristics of wireless IoT network devices.

\section{Mathematical Modeling of the System}
\label{model}

This section presents the mathematical models used in implementing wireless IoT network simulation for FL tasks. To define the energy consumption model and the communication model, the mathematical models presented by \cite{Chen-Walid} were used as a basis, which consider the effects of interference in the uplink and downlink channels in the wireless communication parameters. 

However, this model does not consider attributes for the variation of power and bandwidth in the mathematical formulation of the system model. Therefore, this work uses the parameter discretization technique to adapt the aforementioned mathematical models, where the ranges of values for power adjustment and bandwidth allocation are defined by increments of continuous values. The parameter discretization process is a strategy that simplifies optimization in MILP simulations, reducing the search space and computational complexity in contrast to continuous optimization methods, where decision variables can assume any value within a continuous range.

\subsection{Network Model}
\label{network-model}

In classic ML approaches, all devices send their data to a centralized server to train a conventional model using data from all devices. However, this is not feasible for many wireless IoT network scenarios due to privacy concerns, determining the power level for transmission, and network bandwidth limitations. On the other hand, FL allows multiple devices to collaboratively train a model without sharing data, with the coordination of a parameter aggregator server, where local model training is performed on the device itself \cite{Amannejad}.

Consider a wireless IoT network with a BS directly connected to an FL aggregator server as proposed in the MEC approach, with a set \mbox{$S = \{i_1, i_2, ..., i_N\}$} of $N$ IoT devices before the start of a training round. Each client has a dataset $\mathcal{P}_i$ with $n_i = |\mathcal{P}_i|$ samples stored on their respective local devices. These IoT devices are connected to the BS via a wireless connection and have the ability to collect data and train a local model for a given FL task.

This way, it is possible to train a distributed ML model (e.g., a classifier) without sharing local data, where uplink channels are used to transmit the local parameters of the device models to the BS, while downlink channels are used to transmit global BS parameters to devices. Furthermore, the use of communication resource optimization mechanisms can improve the performance of FL algorithms in wireless IoT networks while also minimizing the energy consumption of the devices.

\subsection{ML Model}
\label{ml-model}

The main idea of FL is to perform the aggregation of a global model over several rounds of communication based on the parameters of the models trained locally on each device. Therefore, the objective of FL is to find the parameters of a global model  $w_{global} = (w^{(1)}, w^{(2)}, ..., w^{(L)})$ that minimize the loss function $f(w_{global})$ using all data $\mathcal{P} = \cup_{i=1}^{N} \mathcal{P}_i$, where each $w_{global}^{(l)} \in \mathcal{R}^{n^{(l)}}$ is the parameter vector of layer $l$ with dimensions $n^{(l)}$ and $L$ is the total number of layers in the model. On the other hand, the local training problem on each device $i$ aims to find the optimal model parameters that minimize its local training loss using the $\mathcal{P}_i$   data. In this way, the FL training process can be expressed as

\vspace{-10pt}
\begin{equation}
\begin{aligned}
    \underset{w_{1}, ..., w_{N}}{\mathrm{min}} f(w_{global}) &\stackrel{\triangle}{=} \underset{w_{1}, ..., w_{N}}{\mathrm{min}} \dfrac{1}{\mathcal{|P|}} \sum_{i=1}^{N} \sum_{j=1}^{n_{i}} f(w_{i}, x_{ij}) \\
    &= \underset{\substack{w_{1}, ..., w_{N}}}{\mathrm{min}} \sum_{i=1}^{N} \dfrac{n_{i}}{m} f(w_{i})
\end{aligned}
\label{eq:modelo-ml}
\end{equation}
\begin{equation}
  \mathrm{s.t.} \hspace{0.5cm} w_{1} = w_{2} = ... = w_{N} = w_{global}\mathrm{,} \tag{1a}
\end{equation}

\vspace{4pt}
\noindent where $f(w_{i}, x_{ij})$ is the loss function that evaluates the performance of the local model $w_{i}$ by observing the output produced by training with $x_{ij}$ data samples, with $m = \sum_{i \in S} n_i$. For example, for supervised tasks, Equation (\ref{eq:modelo-ml}) aims to find the parameters that result in predictions that come as close as possible to the real labels in $x_{ij}$ according to a defined metric by the loss function. The restriction (\ref{eq:modelo-ml}a) guarantees that the models shared between the devices and the BS must be identical and explains the objective of FL in establishing the training of a $w_{global}$ model without data transfer.

The pseudocode in Algorithm \ref{alg-FedAvg} presents the training process of Equation (\ref{eq:modelo-ml}) based on the FedAvg algorithm. Initially, the BS must transmit a global model $w_{0}$ over a wireless link, which must be created or loaded from a pre-trained model. At the beginning of each communication round $t$, the BS must select a random fraction of the $N$ devices, generating a subset $\mathcal{S}_t$ of $m$ devices. Then, the BS uses the downlink channels to transmit the current state of the $w_{t}$ model to each local device.

%\RestyleAlgo{ruled}
\begin{algorithm}
\caption{\textit{Federated Averaging} (FedAvg)}
\label{alg-FedAvg}

 \textbf{Server}:

 \hspace{0.5cm} initialize $w_{0}$

 \hspace{0.5cm} \textbf{for} each round $t=1,2, ...$ \textbf{do}

 \hspace{1cm} $m \gets $max($f_t \cdot \mathcal{N}, 1)$
 
 \hspace{1cm} $\mathcal{S}_{t} \gets $ (random set of $m$ clients)

 \hspace{1cm}  \textbf{for} each client $i \in \mathcal{S}_{t}$ \textbf{in parallel do}

 \hspace{1.5cm} $w_{t+1}^{i} \gets$ ClientUpdate($i$, $w_{t}$)

 \hspace{1cm}  $m_{t} \gets \sum_{i \in \mathcal{S}_{t}} n_i$

  \hspace{1cm} $w_{t+1} \gets \sum_{i \in \mathcal{S}_{t}} \frac{n_{i}}{m_{t}} w_{t+1}^{i}$

\

 \textbf{ClientUpdate($i$, $w$)}: \textit{$\triangleright$ Executed on client $i$}

 \hspace{0.5cm} $\mathcal{B} \gets$ (data $\mathcal{P}_i$ split into batches of size $\mathcal{B})$

 \hspace{0.5cm} \textbf{for} each local epoch $i$ from $1$ to $E$ \textbf{do}

 \hspace{1cm}  \textbf{for} each $b \in \mathcal{B}$ \textbf{do}

\hspace{1.5cm} $w  \gets w - \eta \nabla \ell(w;b)$

 \hspace{0.5cm} \textbf{return} $w$ to server
\end{algorithm}

Each device trains the local model using an optimization algorithm based on Stochastic Gradient Descent (SGD) for each $b \in B$ mini-batches of $\mathcal{P}_{i}$ during $E$ local epochs, where $\nabla \ell$ represents the gradient of $\ell$ in $b$. After the last update $w \gets w - \eta \nabla \ell(w;b)$, the device uses an uplink channel to send the parameters of the local model $w$ to the BS connected to a parameter aggregator server.

This way, the models $w_{t+1}^{i}$ are received and aggregated by the server, generating the current state $w_{t+1}$ of the global model by applying the update $w_{t+1} \gets \sum_{i \in \mathcal{S}{t}} \frac{n{i}}{m_{t}} w_{t+1}^{i}$, where $m_t = \sum_{i \in \mathcal{S}_{t}} n_i$. Aggregation is the weighted average of device parameters, with weights defined based on the amount of data from local devices. The distributed training process can be repeated for several communication rounds until convergence is achieved or the model performance reaches a stopping criterion \cite{Li}.

According to \cite{Chen-Walid}, due to the inherently unreliable nature of the wireless transmission channel, local models received by the BS may be subject to delays or contain incorrect symbols that can affect the regularity of the global model update. This can result in a smaller number of local training and model aggregation iterations impacting the quality of the global FL model.

\subsection{Communication Model}
\label{Communication-Model}

Consider the orthogonal frequency division multiple access (OFDMA) technique for uplink, where each device occupies an RB with the allocation of power and bandwidth discretized with increments of continuous values. Based on \cite{Chen-Walid}, the uplink rate of device \textit{i} transmitting model parameters $w_i$ to the BS can be formulated as

\begin{equation}
    c_{ijkl}^{U} =  \sum_{j=1}^{|B|} \sum_{k=1}^{R} \sum_{l=1}^{|P|} r_{ijkl} B_i^j \mathbb{E} \left( log_{2} \left( 1 + \dfrac{P_{i}^{l}h_{i}}{I_{k} + B_i^jN_{0}} \right) \right) \mathrm{,}
\label{eq:modelo-comunicacao}
\end{equation}

\vspace{10pt}
\noindent
where  $r_{i} = [r_{i111}, ..., r_{iBRP}]$ is an allocation vector with $r_{ijkl} \in {[0,1]}$, $R$ is the number of RBs, and $B$ and $P$ are vectors of discretized elements containing, respectively, the values for bandwidth and power allocation. Furthermore, $\sum_{j=1}^{|B|}\sum_{k=1}^{R}\sum_{l=1}^{|P|} r_{ijkl}=1$, with $r_{ijkl}=1$ indicating that the uplink rate of device $i$ is $c_{ijkl}^{U}$ using bandwidth $B_i^j$ on RB $k$ with power $P_{i}^{l}$. The channel gain between device $i$ and the BS is given by $h_{i} = o_{i}d_{i}^{-\alpha}$, where $d_{i}$ is the distance between device $i$ and the BS, $o_{i}$ is the Rayleigh fading parameter, and $\alpha$ is an exponent that affects how the channel gain varies with distance. $\mathbb{E}(.)$ is the expected data rate with respect to $h_{i}$, $N_{0}$ is the noise power spectral density, and $I_{k}$ is the interference to RB $k$ caused by other devices.

The transmit power of the BS is generally much higher than the power of the devices. Therefore, the entire bandwidth of the downlink can be used for the transmission of the global model. Thus, the downlink data rate achieved by the BS when transmitting the global model parameters to each device is given by

\begin{equation}
    c_{i}^{D} = B^{D} \mathbb{E} \left( log_{2} \left( 1 + \dfrac{P_{B}h_{i}}{I^{D} + B^{D}N_{0}} \right) \right) \mathrm{,}
\label{eq:modelo-comunicacao}
\end{equation}

\noindent
where $B^{D}$ is the bandwidth used by the BS to transmit the global model to each device, $P_{B}$ is the transmission power of the BS, and $I^{D}$ is the interference caused by other BSs that do not participate in the FL task.

To formulate the transmission delay, it is assumed that FL models are transmitted via a single packet. Thus, the transmission delay between a device $i$ and the BS at uplink and downlink can be respectively formulated as

\vspace{10pt}
\begin{minipage}{0.22\textwidth}
\begin{equation}
    l_{ijkl}^{U} = \dfrac{S_{pkt}^{U}}{c_{ijkl}^{U}} \mathrm{,}
\label{eq:modelo-delay-uplink}
\end{equation}
\end{minipage}
\hfill
\begin{minipage}{0.22\textwidth}
\begin{equation}
    l_{k}^{D} = \dfrac{S_{pkt}^{D}}{c_{i}^{D}} \mathrm{,}
\label{eq:modelo-delay-downlink}
\end{equation}
\end{minipage}
\vspace{5pt}

\noindent
where $S_{pkt}^{U}$ is the size of the uplink packet, that is, the number of bits that the devices need to transmit to the BS through the wireless link, considering the size of the parameters of the local models plus other information, such as the accuracy of training on the device and the value of the local loss function. Likewise, $S_{pkt}^{D}$ is denoted as the size of the downlink packet, that is, the number of bits that the BS needs to transmit to the devices over the wireless link, considering the size of the global model parameters plus other information that may indicate some instruction from the server to the devices.

Thus, in one round of communication, the total transmission delay between a device $i$ and the BS is given by

\begin{equation}
    l_{ijkl}^{R} = l_{ijkl}^{U} + l_{ijkl}^{D}
\label{eq:modelo-energia}
\end{equation}

\noindent
where $l_{ijkl}^{R}$ is equivalent to the sum of the uplink and downlink transmission delays.

To simulate the effect of packet transmission errors, it is considered that the BS will not ask devices to resend models when they are received with errors. Therefore, whenever a packet containing a received local model contains errors, the BS will not use it to update the global model. In this case, based on \cite{Chen-Walid}, the packet transmission error rate of the uplink from device \textit{i} is given by

\begin{equation}
    q_{ijkl}^{U} = \sum_{j=1}^{B} \sum_{k=1}^{R} \sum_{l=1}^{P} r_{ijkl} \mathbb{E} \left( 1 - \mathrm{exp} \left( - \dfrac{m (I_{k} + B_i^jN_{0})}{P_{i}^{l}h_{i}} \right) \right) \mathrm{,}
\label{eq:modelo-erro-pacote}
\end{equation}

\vspace{10pt}
\noindent
where $\mathbb{E}(.)$ is the expected packet error rate considering $h_{i}$ in RB $k$, with $m$ being a threshold (\textit{waterfall threshold}) that defines transmission quality.

Considering that the successful transmission of a packet from the devices to the BS without errors is the complement of the uplink packet error rate, the probability of successful transmission of the local model to the BS can be formulated as

\begin{equation}
    p_{ijkl}^{U} = 1 - q_{ijkl}^{U} \mathrm{.}
\label{eq:modelo-prob-sucesso}
\end{equation}

\vspace{4pt}
The estimation of $q_{ijkl}^{U}$ or $p_{ijkl}^{U}$ in packet transmission is useful for evaluating the efficiency of allocating communication resources to devices participating in an FL task. For example, if the probability of success is significant, this may indicate that the transmission of a local model will have a greater chance of being received by the BS and added to the global model by the parameter server. Finally, the probability of successful transmission of the local model to the BS is denoted considering compliance with a policy as follows

\begin{equation}
\label{eq:ajuste-prob}
    p\gamma_{ijkl}^{U}=    \begin{cases}
        p_{ijkl}^{U} & \text{if} \ (l_{ijkl}^{U} \leq \gamma_{T}  \text{ \& }  e_{ijkl}^U \leq \gamma_{E} 
        \text{ \& }  q_{ijkl}^{U} \leq \gamma_{Q}), \\
        0 & \text{otherwise} \mathrm{,}
    \end{cases}
\end{equation}

\noindent
where $\gamma_{T}$ defines the delay requirement, $\gamma_{E}$ defines the energy consumption requirement, and $\gamma_{Q}$ defines the minimum packet transmission error rate requirement for sending the model $w_i$ in a communication round. In case of transmission failure, model $w_i$ is not received by the BS and therefore does not contribute to the aggregation of model $w_{global}$.

\subsection{Energy Consumption Model}
\label{Energy-Model}

The energy consumption model incorporates the energy required for local model training and transmission. The energy demand of the BS is not considered, since it normally has a continuous energy supply. Based on \cite{Chen-Walid}, the device power consumption model for training and transmission of the $w_i$ model can be formulated as

\vspace{4pt}
\begin{minipage}{0.22\textwidth}
\begin{equation}
    e_{ijkl}^{T} = \zeta \omega_{i} \vartheta^{2} Z(w_{i}) \mathrm{,}
\label{eq:modelo-energia-treinamento}
\end{equation}
\end{minipage}
\hfill
\begin{minipage}{0.22\textwidth}
\begin{equation}
    e_{ijkl}^{U} = P_{i}^{l} l_{ijkl}^{U} \mathrm{,}
\label{eq:modelo-energia-upload}
\end{equation}
\end{minipage}
\vspace{10pt}

\noindent
where $\vartheta$, $\omega_{k}$, and $\zeta$ refer, respectively, to the clock frequency, the number of cycles of the central processing unit, and the energy consumption coefficient of each device. In Equation \ref{eq:modelo-energia-upload}, $P_{i}^{l}$ defines the power allocated to the device $i$ multiplied by the value of $l_{ijkl}^{U}$ which defines the transmission delay between device $i$ and the BS on the uplink using bandwidth $B_i^j$ on RB $k$ with power $P_{i}^{l}$.

Thus, in one round of communication, the computation of the energy consumed by device $i$ is given by

\begin{equation}
    e_{ijkl}^{R} = e_{ijkl}^{T} + e_{ijkl}^{U}
\label{eq:modelo-energia}
\end{equation}

\noindent
where $e_{ijkl}^{R}$ is equivalent to the sum of the energy consumption for training and transmission of the model $w_i$.

\section{Algorithm Formulation}
\label{problem-formulation}

To solve the general problem of Equation (\ref{eq:modelo-ml}), \mbox{FL-E2WS} proposes its division into two subproblems. The first, responsible for device selection, is presented in Section \ref{selection-devices}, and the second, responsible for scheduling communication resources, is presented in Section \ref{allocation-rbs}. Finally, Section \ref{algorithm-proposal} presents the FL-E2WS algorithm.

\subsection{Device Selection}
\label{selection-devices}

Although FL may involve hundreds or millions of clients, due to the high overhead of aggregating updates, in practice, only a small fraction of clients participate in each round of training \cite{CLiZeng}. In an FL task, each device contributes a local model trained with data covering different characteristics, contexts, and variations of the problem. 

Thus, by directing the strategy towards local training that covers more data, FL-E2WS is expected to improve the representativeness of locally trained models. Therefore, by denoting $S_{t}$ as a fraction $f_{t}$ of $N$ devices and $b$ as a binary vector indicating the selection of $n_{p}$ devices, the problem of maximizing the quantity of data from $n_{p} \leq |S_{t}|$ devices can be expressed as

% \vspace{4pt}
\begin{equation}
    \mathrm{max} \sum_{i \in S_{t}} |\mathcal{P}_{i}|b_{i}  \mathrm{,}  
\label{eq:selecao-disp}
\end{equation}

\vspace{6pt}
\begin{equation}
    \hspace{0.2cm} \mathrm{s.t.} \hspace{0.2cm} \sum_{i=1}^{|S_{t}|} b_{i} = n_{p} \mathrm{,} \tag{\ref{eq:selecao-disp}a} \\
\end{equation}

\vspace{2pt}
\begin{equation}
      \hspace{0.8cm}
     b_{i} \in \{0,1\}   \mathrm{,} \tag{\ref{eq:selecao-disp}b} \\
\end{equation}

\vspace{5pt}
\noindent
where $|\mathcal{P}i|$ is the number of data samples from device \mbox{$i \in S{t}$} and $n_{p}$ is the partial number of devices selected for the resource scaling step of communication. The constraint (\ref{eq:selecao-disp}a) guarantees that the number of selected devices is equal to $n_{p}$, and the constraint (\ref{eq:selecao-disp}b) indicates that $b$ is a binary vector, where $b_{i} = 1$ indicates that the device was selected from the subset $S_{t}$ and $b_{i} = 0$ indicates otherwise.

\subsection{Communication Resource Scheduling}
\label{allocation-rbs}

This work proposes the scheduling of communication resources with the objective of minimizing the energy consumption of transmitting the local models of the devices to the BS while maximizing the number of devices selected for each round of communication.

By denoting the binary matrix $c_{ijkl}$ as an indicator of the assignment of device $i$ with bandwidth $B_i^j$ in RB $k$ with power $P_{i}^{l}$, the problem of scheduling communication resources can be formulated as

%\vspace{-10pt}
\begin{equation}
\begin{aligned}
    \mathrm{min} \sum_{i=1}^{n_{p}} \sum_{j=1}^{|B|} \sum_{k=1}^{R} \sum_{l=1}^{|P|} P_{i}^{l} \times l_{ijkl}^{U} \times c_{ijkl} - \lambda \sum  c_{ijkl}
   \mathrm{,}  
\end{aligned}
\label{eq:escalonamento-recursos}
\end{equation}
\vspace{-2pt}
\begin{equation}
 \mathrm{s.t.}\hspace{0.5cm} \sum_{i=1}^{n_{p}} \left(\sum_{j=1}^{|B|} \sum_{k=1}^{R} \sum_{l=1}^{|P|} c_{ijkl}\right) = 1 \mathrm{,} \tag{\ref{eq:escalonamento-recursos}a} \\
\end{equation}
\begin{equation}
 \hspace{0.8cm}\hspace{0.5cm}  \sum_{k=1}^{R} \left(\sum_{j=1}^{|B|} \sum_{i=1}^{n_{p}} \sum_{l=1}^{|P|} c_{ijkl}\right) = 1 \mathrm{,} \tag{\ref{eq:escalonamento-recursos}b} \\
\end{equation}
\begin{equation}
  \hspace{1cm} l_{ijkl}^{U} \leq \gamma_{T}  \mathrm{,} \tag{\ref{eq:escalonamento-recursos}c} \\
\end{equation}
\begin{equation}
  \hspace{1cm} e_{ijkl}^{U} \leq \gamma_{E}  \mathrm{,} \tag{\ref{eq:escalonamento-recursos}d} \\
\end{equation}
\begin{equation}
  \hspace{1cm} q_{ijkl}^{U} \leq \gamma_{Q} \mathrm{,}  \tag{\ref{eq:escalonamento-recursos}e} \\
\end{equation}
\begin{equation}
  \hspace{1cm} P_i^{min} \leq P_{i}^{l} \leq P_i^{max} \mathrm{,}  \tag{\ref{eq:escalonamento-recursos}f} \\
\end{equation}
\begin{equation}
  \hspace{1cm} \sum B_i^j \leq B^T   \mathrm{,} \tag{\ref{eq:escalonamento-recursos}g} \\
\end{equation}
\begin{equation}
  \hspace{1.3cm} \sum c_{ijkl} \leq n_{f}  \mathrm{,} \tag{\ref{eq:escalonamento-recursos}h} \\
\end{equation}
\begin{equation}
 \hspace{1.5cm} c_{ijkl} \in \{0,1\}   \mathrm{,} \tag{\ref{eq:escalonamento-recursos}i} \\
\end{equation}

\noindent
where $\lambda$ is a weight that controls the importance of the objective of maximizing the number of devices, with $n_{f} \leq n_{p}$ being the maximum number of devices that should be selected for the next round of communication. The constraint (\ref{eq:escalonamento-recursos}a) guarantees that each device is allocated a single bandwidth $B_i^j$ and power $P_{i}^{l}$ assigned to exactly one RB $k$, and the constraint (\ref{eq:escalonamento-recursos}b) guarantees that each RB is allocated to a single device and power. The constraints (\ref{eq:escalonamento-recursos}c), (\ref{eq:escalonamento-recursos}d), and (\ref{eq:escalonamento-recursos}e) establish, respectively, compliance with the delay requirement, energy consumption, and packet transmission error rate for sending model $w_i$ in one round of communication. The constraint (\ref{eq:escalonamento-recursos}f) establishes the minimum and maximum power of the devices, the constraint (\ref{eq:escalonamento-recursos}g) defines that the total sum of $B_i^j$ must not exceed $B^T$, which defines the total bandwidth budget. The constraint (\ref{eq:escalonamento-recursos}h) guarantees that the final number of selected devices is at most $n_{f}$, and the constraint (\ref{eq:escalonamento-recursos}i) defines that $c$ is a binary matrix, where $c_{ijkl} = 1$ indicates that RB $k$ has been allocated to device $i$ with bandwidth $B_i^j$ and power $P_{i}^{l}$, and $c_{ijkl} = 0$ indicating otherwise.

It is observed that the formulation of the objective function of Equation (\ref{eq:escalonamento-recursos}) has two functions that simultaneously seek to optimize different objectives. The first seeks to minimize the energy consumption of devices, and the second seeks to maximize the number of selected devices that should consume energy for training and transmitting local models. In this case, the definition of different objectives is combined in the objective function, with the second function having its negative contribution and using the weight $\lambda$ to indicate the importance of maximizing the number of devices selected for the next round of communication.

Therefore, the $\lambda$ adjustment balances the relative importance of the two objectives. If $\lambda$ is large, maximizing the number of selected devices will have a significant influence on the objective function. If $\lambda$ is small, the objective function will prioritize minimizing the total energy cost, even if the maximum number of devices is not selected for the next round of communication.

\subsection{FL-E2WS Algorithm}
\label{algorithm-proposal}

The Algorithm \ref{alg} presents the FL-E2WS strategy in wireless IoT networks based on Equations (\ref{eq:selecao-disp}) and (\ref{eq:escalonamento-recursos}). In the device selection stage, it is observed that the aggregator server selects $n_p$ devices, maximizing the amount of local data. Then, communication resources are scheduled for $n_f$ devices, where the binary matrix $c$ defines the devices selected for the next round of communication, considering the ideal bandwidth and power for transmitting the local models under a given device selection and allocation of uplink RBs, meeting the delay requirement $\gamma_{T}$, the energy consumption requirement $\gamma_{E}$, and the packet error rate requirement $\gamma_{Q}$.

\begin{algorithm}
\caption{\textit{FL-E2WS for FL in Wireless IoT Networks}}
\label{alg}

\textbf{Server}:

 \hspace{0.5cm} initialize $w_{0}$

 \hspace{0.5cm} \textbf{for} each round $t=1,2, ...$ \textbf{do}
 
 \hspace{1cm} $\mathcal{S}_{t} \gets $ (random set of $max(f_{t} \cdot \mathcal{N}, 1)$ devices)

\hspace{1cm} \textit{$\triangleright$\small Maximize the amount of data for $n_p$ devices
 \small} 

\hspace{1cm} Define $b$ with $n_{p} \leq |S_{t}|$ devices using Equation (\ref{eq:selecao-disp})

\hspace{1cm} \textit{$\triangleright$\small Schedule resources to $n_f$ devices with $\gamma_{T}$, $\gamma_{E}$ and $\gamma_{Q}$\small}

\hspace{1cm} Define $c$ with $n_{f} \leq n_{p}$ devices using Equation (\ref{eq:escalonamento-recursos})

 \hspace{1cm}  \textbf{for} each client $i \in \mathcal{S}_{t}$ \textbf{in parallel do}

 \hspace{1.5cm} $w_{t+1}^{i} \gets$ DeviceSelected($i$, $w_{t}$)

 \hspace{1cm}  $m_{t} \gets \sum_{i \in \mathcal{S}_{t}} n_i$

  \hspace{1cm} $w_{t+1} \gets \sum_{i \in \mathcal{S}_{t}} \frac{n_{i}}{m_{t}} w_{t+1}^{i}$

\

 \textbf{DeviceSelected($i$, $w$)}: \textit{$\triangleright$ For each device
 $i$}

 \hspace{0.5cm} $\mathcal{B} \gets$ (data $\mathcal{P}_i$ split into batches of size $\mathcal{B})$

 \hspace{0.5cm} \textbf{for} each local epoch $k$ from $1$ to $E$ \textbf{do}

 \hspace{1cm}  \textbf{for} each $b \in \mathcal{B}$ \textbf{do}

\hspace{1.5cm} $w  \gets w - \eta \nabla \ell(w;b)$

 \hspace{0.5cm} \textbf{return} $w$ to server
\end{algorithm}

\section{Performance Evaluation}
\label{results}

\begin{figure*}[!t]
\centering
\subfloat[]{\includegraphics[width=0.3\textwidth]{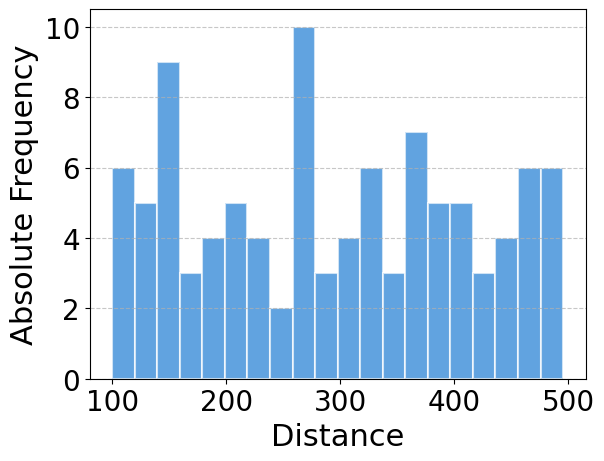} 
\label{abs}}
\hfil
\subfloat[]{\includegraphics[width=0.3\textwidth]{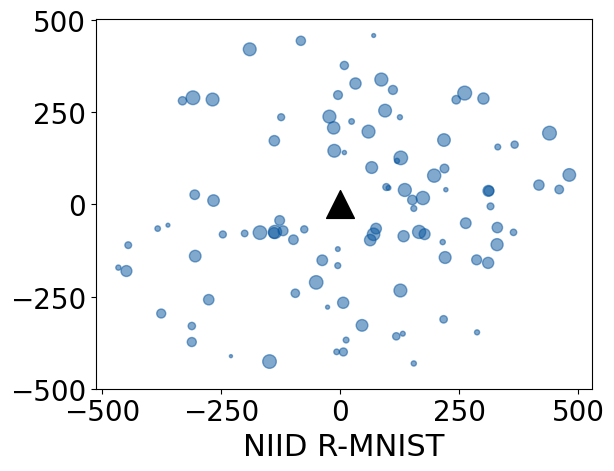}
\label{distribuicao-mnist}}
\hfil
\subfloat[]{\includegraphics[width=0.3\textwidth]{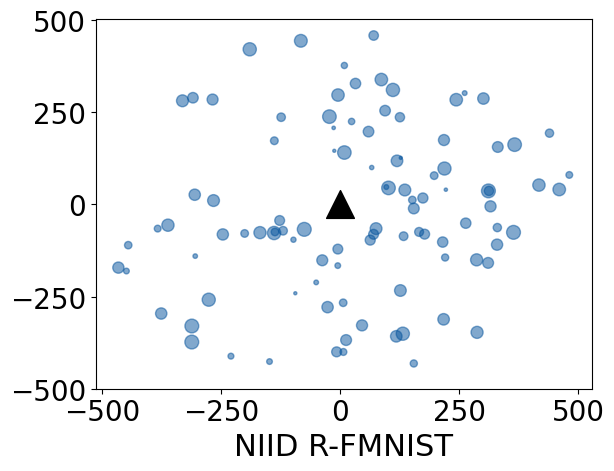}
\label{distribuicao-fashion}}
\caption{Distance from devices to the BS in meters. (a) Absolute Frequency. (b) NIID R-MNIST dataset. (c) NIID R-FMNIST dataset. In (b) and (c), each device is represented by a circle, with the circle's size being proportional to the amount of data.}
\label{fig:distancia}
\end{figure*}

Consider a wireless IoT network serving a circular area with a radius $r$ of $500$ meters with a BS in the center. The BS is directly associated with an aggregator server, where there are $N=100$ devices connected for an FL task. As shown in Figure \ref{fig:distancia}, the devices are distributed uniformly and randomly at distances between $100$ and $500$ meters from the BS.

The uplink bandwidth of each RB is limited by discretized values generated by an arithmetic progression in the range of $[1, 2]$ MHz with increments and median values of $1 \times 10^{-1}$ MHz and $1.5$ MHz. Similarly, the transmission power of the devices is limited by discretized values generated by an arithmetic progression in the range of $[0.008, 0.012]$ W with increments and median values of $2.5 \times 10^{-4}$ W and $0.01$ W. The total bandwidth budget for the uplink is equivalent to the value of $n_f \times 1.5$ MHz, where $n_f$ defines the maximum number of devices participating in each round of communication. Furthermore, each RB includes the attribution of a distinct and incremental interference. The downlink bandwidth is $20$ MHz and the BS transmit power is $1$ W. The $h_{k}$ modeling incorporates a fading effect that indicates that the channel gain decreases as the distance of the devices from the BS increases. In Table \ref{tab:parametros}, other simulation parameters are presented.

\begin{table}[h]
    \begin{center}   
    \caption{Simulation Parameters
 \cite{Chen-Walid}}
\begin{tabular}{|c|c|c|c|}
\hline

\textbf{Parameter} & \textbf{Value} & \textbf{Parameter} & \textbf{Value}   \\
\hline
\hline
$\alpha$ & $2$ & $\vartheta$ & $10^{9}$ \\
\hline
$N_0$ & $-174$ dBm/Hz & $\zeta$ & $10^{-27}$ \\
\hline
$m$ & $0.023$ dB & $\omega_{i}$ & $40$ \\
\hline

\hline
   \end{tabular}
     \label{tab:parametros}
    \end{center}   
\end{table}

FL tasks consider image classification problems using the basis of benchmark datasets used in FL research, called MNIST \cite{mnist} and Fashion-MNIST \cite{Fashion-MNIST}, according to the works of \cite{Beutel}, \cite{McMahan}, \cite{Chen-Walid}, \cite{Zhu-Guangxu}, \cite{Chen-Hao},  \cite{AnycostFL}, \cite{Peng}, \cite{Amannejad}, \cite{Zhao}, \cite{RL-Zhou} and \cite{Bhatti}. MNIST consists of handwritten digits distributed across $10$ classes, where each class is associated with a digit from $0$ to $9$. On the other hand, Fashion-MNIST diversifies the context by presenting $10$ classes representing different clothing categories such as t-shirts, pants, shoes, etc.

Unlike the references cited, this work sought to make device data heterogeneous as an inherent characteristic of wireless IoT networks. By using variations of MNIST and Fashion-MNIST containing statistical heterogeneity in the data distribution in a non-Independent and Identically Distributed (non-IID) manner, a challenge was introduced for optimizing the accuracies of local models and aggregation of the global model. This avoided trivial convergence to values close to totality and allowed a more notable observation about the performance of the analyzed algorithms.

Thus, the MNIST and Fashion-MNIST datasets were divided into $10$ subsets with samples from the same label, using $75\%$ of the samples for the training set and $25\%$ for the test set. Then, each device received a training and testing partition, where: $90\%$ of the samples belong to the same label and the remaining $10\%$ belong equally to the other labels; each image is rotated up to $45^{\circ}$ clockwise or counterclockwise; the final amount of data is given by a factor between $[0.25, 1]$ of the initial partition. In this work, the MNIST and Fashion-MNIST partitions were named, respectively, as NIID R-MNIST and NIID R-FMNIST.

\begin{figure}[!t]
\centering
\subfloat[]{\includegraphics[width=0.38\textwidth]{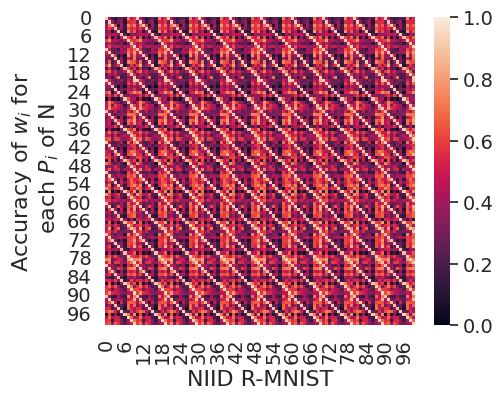}
\label{mnist-distribuicao-de-dados}
}
\hfil
\subfloat[]{\includegraphics[width=0.38\textwidth]{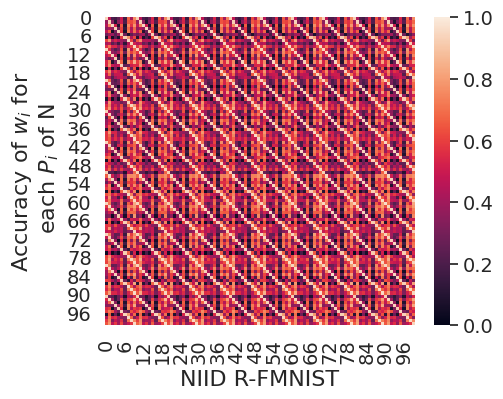}
\label{fmnist-distribuicao-de-dados}
}
\caption{Accuracy of $w_i$ using local data from other devices. (a) NIID R-MNIST dataset. (b) NIID R-FMNIST dataset.}
\label{fig:ds-distribuicao-dados}
\end{figure}

Figures \ref{fig:distancia} and \ref{fig:ds-distribuicao-dados} present an intuition about the non-IID distribution of data samples from each device for the NIID R-MNIST and NIID R-FMNIST datasets. In Figure \ref{fig:distancia}, each device is represented by a circle, where the size of the circle is proportional to the amount of data. On the other hand, Figure \ref{fig:ds-distribuicao-dados} presents the evaluation of the $w_i$ model using local data from the device and test data from the other $\mathcal{S}$ devices for each dataset. Each element of the main diagonal represents the accuracy of the model $w_i$ trained with its local data set. The other elements of each line indicate the accuracy of $w_i$ tested with local data from other users. Therefore, it is possible to observe that the $w_i$ model achieves good performance when considering local training with data from the device itself. However, the $w_i$ model has difficulty generalizing to other datasets, except when repeating the data distribution pattern for other devices.

\begin{comment}
\begin{figure}[ht]
    \centering
    \hspace{-0.10cm}
    \begin{minipage}[b]{0.22\textwidth}
        \centering
        \subfloat[]{\includegraphics[width=1\textwidth]{fig/mnist-distribuicao-de-dados.png}%        
}       
    \end{minipage}    
    \hspace{0.1cm} 
    \begin{minipage}[b]{0.22\textwidth}
        \centering
        \subfloat[]{\includegraphics[width=1\textwidth]{fig/fmnist-distribuicao-de-dados.png}%
}     
    \end{minipage}
    \caption{Accuracy of $w_i$ using local data from other devices}
    \label{fig:ds-distribuicao-dados}
\end{figure}
\end{comment}

For the classification task of the NIID R-MNIST dataset, we consider a Multi-Layer Perceptron (MLP) neural network architecture composed of a single hidden layer with 128 neurons, followed by the ReLU activation and a softmax output layer. Furthermore, for the task of classifying the NIID R-FMNIST dataset, a neural network architecture of the Convolutional Neural Network (CNN) type is adopted, which incorporates two convolutional layers of dimensions $3 \times 3$, the first with $16$ and the second with $32$ convolutional filters, where each convolutional layer is followed by a MaxPooling layer of dimensions $2 \times 2$. The architecture is finalized by applying the ReLU activation function, followed by a softmax output layer. The MLP and CNN architectures have respectively $101,770$ and $206,922$ parameters, as well as using the ADAM optimizer (Adaptive Moment Estimation) and the Sparse Categorical Crossentropy loss function.

It is noteworthy that it is possible to use different architectures of deeper neural networks in order to achieve better model performance. However, according to \cite{Zhao}, even if the model's accuracy does not reach the state of the art, this may still be sufficient for evaluating the behavior of FL strategies in different scenarios.

\subsection{Simulation Results}
\label{results-simulation}

This section discusses the results of FL-E2WS compared to the FedAvg and POC algorithms previously discussed in related works. The FedAvg algorithm randomly determines device selection and communication resource allocation. On the other hand, POC selects devices based on the highest values of the local loss function, which can be previously sent by devices along with local models without any knowledge of the wireless network. Furthermore, to ensure fairness in the comparison with FL-E2WS, variations of FedAvg and POC were implemented with fixed network bandwidth and transmission power allocation, referred to in this work as FedAvg-w$_{Opt}$ and POC-w$_{Opt}$, incorporating the knowledge of the wireless network and using the MILP algorithm of \mbox{FL-E2WS} for assigning uplink RBs.

Due to the inherent stochastic nature of the simulation, each FL algorithm was run $15$ times for each dataset in order to obtain statistics related to average performance. The FL algorithms were executed for $200$ communication rounds, using the number of uplink RBs proportional to the value of $n_f$ that defines the maximum number of devices participating in each communication round. In this case, FedAvg and \mbox{FedAvg-w$_{Opt}$} used the value of $n_f$, and the other strategies used the value of $n_p = n_f \times 1.5$ to define the initial subset of devices for applying the selection strategies for the final number of devices. Each model was trained for one local training epoch. In all strategies, at most $n_f$ devices were selected in each communication round after allocating communication resources in accordance with the policy that establishes delay, energy consumption, and packet error rate requirements.

Considering that MLP and CNN have a different number of parameters, two policies were defined for the evaluation of \mbox{FL-E2WS}, ensuring that any device could transmit its model in at least one uplink RB with $\gamma_{Q}=0.3$ as the minimum packet transmission error rate requirement. For NIID R-MNIST with MLP, $\gamma_{T}=200$ ms was used as the delay requirement and $\gamma_{E}=0.0025$ J as the energy consumption requirement. For NIID R-FMNIST with CNN, were used $\gamma_{T}=400$ ms and $\gamma_{E}=0.005$ J.

\begin{figure*}[!t]
\centering
\subfloat[]{\includegraphics[width=0.42\textwidth]{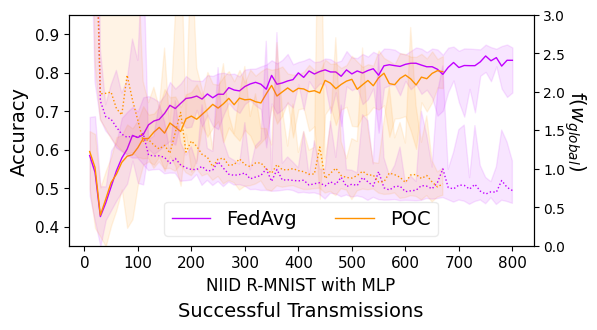}
\label{mnist-sem-otimizacao}}
\hfil
\subfloat[]{\includegraphics[width=0.42\textwidth]{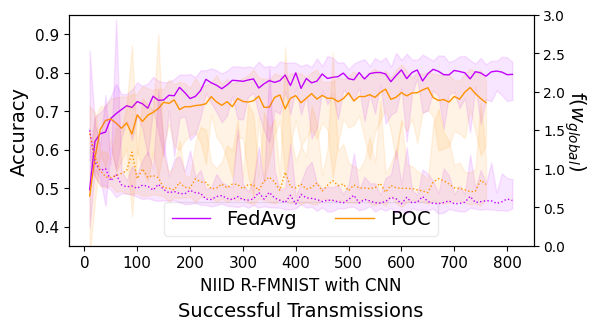}
\label{fmnist-sem-otimizacao}}
\vfil
\subfloat[]{\includegraphics[width=0.42\textwidth]{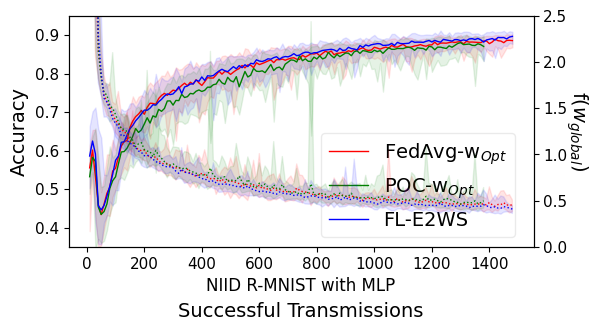}
\label{mnist-com-otimizacao}}
\hfil
\subfloat[]{\includegraphics[width=0.42\textwidth]{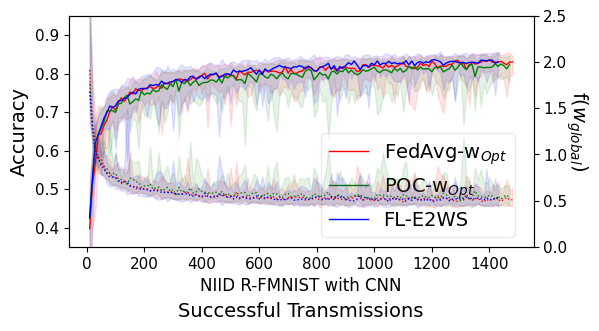}
\label{fmnist-com-otimizacao}}
\caption{Evolution of accuracy and $f(w_{global})$ of FL algorithms with NIID R-MNIST and NIID R-FMNIST datasets. In (a) and (b), it is the FL algorithms that do not know the transmission channel. FedAvg and POC are unable to successfully transmit a significant number of local models. In (c) and (d), it is the FL algorithms that incorporate knowledge of the transmission channel enabling them to successfully transmit a significantly greater number of local models. FedAvg-w$_{Opt}$ and POC-w$_{Opt}$ incorporate knowledge of the transmission channel using the FL-E2WS algorithm operating with fixed bandwidth and transmission power. FL-E2WS prioritizes minimizing the energy consumption of devices by optimizing the allocation of uplink RBs, bandwidth and transmission power.}
\label{fig_mnist_acuracia}
\end{figure*}

Figures \ref{mnist-sem-otimizacao} and \ref{fmnist-sem-otimizacao} show the evolution of accuracies and $f(w_{global})$ of the FedAvg and POC algorithms considering the number of successful local model transmissions using the mentioned datasets, where at most $n_f=10$ devices are selected in each communication round. The solid curves represent the average behavior, and the shaded regions correspond to the minimum and maximum values over the $15$ executions to obtain statistics related to the performance of each algorithm.

The results show that the accuracy of FL algorithms increase as the number of local model transmissions increases. In this case, progressively increasing the sharing of local models during training rounds allows the global model to benefit from the information acquired in the devices data patterns. Similarly, as the number of transmissions increases, the value of $f(w_{global})$ of the algorithms decreases. Furthermore, it is noted that the FedAvg and POC algorithms are unable to successfully transmit a significant number of local models due to the lack of knowledge of the conditions of the transmission channel. In this case, when few models are delivered to the aggregator server, there is a decrease in the accuracy of the global model due to the loss of representativeness of data from devices inherent to the transmitted models.

Figures \ref{mnist-com-otimizacao} and \ref{fmnist-com-otimizacao} present the evolution of accuracies and $f(w_{global})$ of the algorithms that incorporate knowledge of the transmission channel, considering the number of successful model transmissions with $n_f=10$. The variations of FedAvg and POC, referred to in this work as FedAvg-w$_{Opt}$ and \mbox{POC-w$_{Opt}$}, use fixed bandwidth and transmission power, respectively, with values of $1$ MHz and $0.01$ W.

When incorporating the knowledge of the wireless IoT network using the MILP algorithm of \mbox{FL-E2WS} for the assignment of the uplink RBs, FedAvg-w$_{Opt}$ and POC-w$_{Opt}$ algorithms were able to successfully transmit a significantly greater number of local models compared to their original versions that are unaware of the conditions of the transmission channel. This resulted in better accuracy performance of the global model.

It is noteworthy that FL-E2WS prioritizes minimizing the energy consumption of devices, and yet, it demonstrated similar performance to other algorithms that operate with fixed bandwidth and transmission power. In addition to the allocation of uplink RBs, the performance of FL-E2WS is characterized by the optimization and allocation of bandwidth and transmission power, allowing an increase in the number of transmissions of local models while also considering the minimization of energy consumption. At the same time, the targeted choice of devices for local training, covering a broader set of data, improves the representativeness of the trained models, contributing to the ability of FL-E2WS to generalize the global model for correct predictions.

\begin{figure*}[!t]
\centering
\subfloat[]{\includegraphics[width=0.42\textwidth]{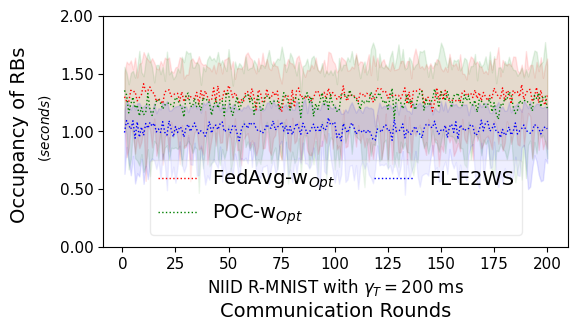}
\label{mnist-ocupacao-rbs}}
\hfil
\subfloat[]{\includegraphics[width=0.42\textwidth]{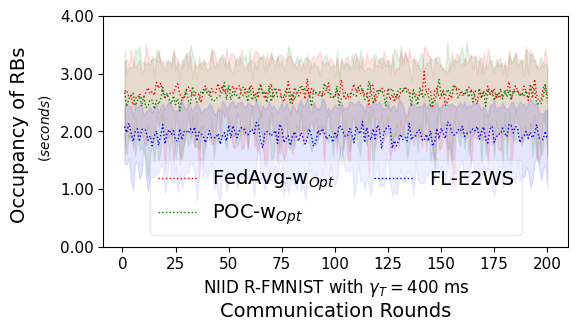}
\label{fmnist-ocupacao-rbs}}
\caption{Sum of the RBs occupation times in each communication round. FL-E2WS enhances communication resource utilization by minimizing the models transmission time of local models to the BS.}
\label{ocupacao-rbs}
\end{figure*}

In Figures \ref{mnist-ocupacao-rbs} and \ref{fmnist-ocupacao-rbs}, the sum of the RBs occupation times in each communication round is presented with $n_f=10$. In this case, FL-E2WS improves the use of communication resources by reducing the transmission time of the models. Therefore, FL-E2WS contributes to minimizing energy consumption per unit of transmitted data, both by reducing the transmission time of the models and by reducing the operating time of the devices.

\begin{figure}[!t]
\centering
\subfloat[]{\includegraphics[width=0.48\textwidth]{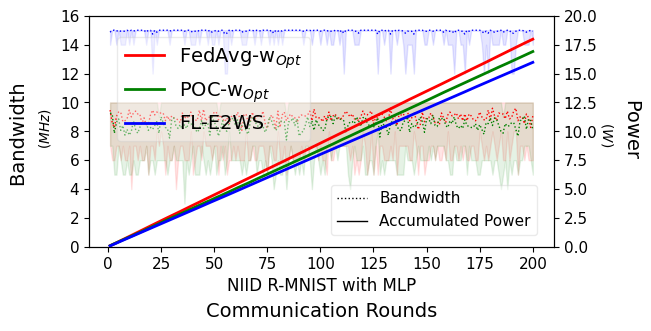}
\label{mnist-power}
}
\hfil
\subfloat[]{\includegraphics[width=0.48\textwidth]{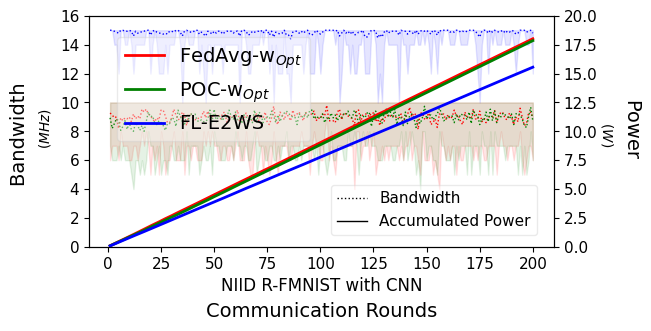}
\label{fmnist-power}
}
\caption{Bandwidth and power allocation. FL-E2WS dynamically allocates communication resources based on the demand of the devices selected for each round in order to maximize energy efficiency.}
\label{fig:power}
\end{figure}

Figures \ref{mnist-power} and \ref{fmnist-power} show the sum of bandwidth and the evolution of the accumulated power allocated to the devices in each communication round with $n_f=10$. \mbox{FL-E2WS} dynamically allocates communication resources based on the demand of the devices selected for each round. At the same time, FL-E2WS maximizes the utilization of the total bandwidth budget and minimizes the power allocated to devices. Dynamic allocation allows resources to be automatically adjusted to best meet the policy that defines the delay requirement $\gamma_{T}$, the energy consumption requirement $\gamma_{E}$, and the minimum packet transmission error rate requirement $\gamma_{Q}$.

The results in Figures \ref{mnist-com-otimizacao}, \ref{fmnist-com-otimizacao}, \ref{ocupacao-rbs} and \ref{fig:power} are in accordance with the formulation of the objective function in Equation (\ref{eq:escalonamento-recursos}) from FL-E2WS. Therefore, this means that FL-E2WS is able to minimize power and transmission time while maximizing the number of devices selected in each communication round. This results in lower energy consumption during the model transmission process, which is particularly important for wireless IoT networks, where energy efficiency is crucial due to device power constraints.

\begin{figure*}[!t]
\centering
\subfloat[]{\includegraphics[width=0.42\textwidth]{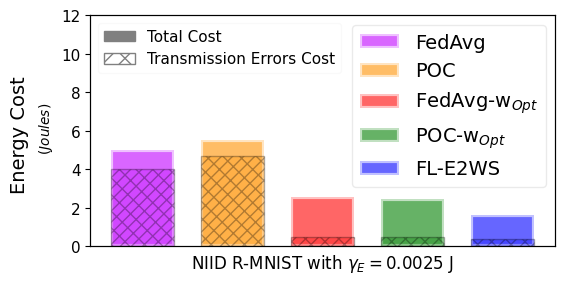}
\label{mnist-energia}}
\hfil
\subfloat[]{\includegraphics[width=0.42\textwidth]{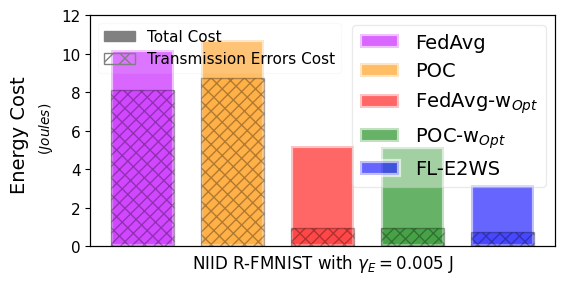}
\label{fmnist-energia}}
\caption{The FedAVG and POC algorithms that are unaware of the transmission channel consume most of the energy resources in training and attempting to transmit models, which results in transmission errors. FL-E2WS demonstrates efficiency in allocating resources more assertively for the use of energy resources, being more efficient both in reducing total energy consumption and in minimizing energy consumption associated with transmission errors.}
\label{fig:custo-energetico}
\end{figure*}

In this context, the bars with solid filling in Figures \ref{mnist-energia} and \ref{fmnist-energia} present the total energy cost, and the bars with crossed-out filling present the energy cost of transmission errors for devices with $n_f=10$. Due to the lack of knowledge of the conditions of the transmission channel, the FedAvg and POC algorithms consume most of the energy resources in training and attempting to transmit models, which results in transmission errors. On the other hand, the \mbox{FedAvg-w$_{Opt}$}, POC-w$_{Opt}$, and FL-E2WS algorithms demonstrate efficiency in allocating resources in a more assertive way for the use of the energy resources of devices for successful model training and transmissions. FL-E2WS is more efficient in both reducing total energy consumption and minimizing energy consumption associated with transmission errors.

\begin{figure}[!t]
\centering
\subfloat[]{\includegraphics[width=0.45\textwidth]{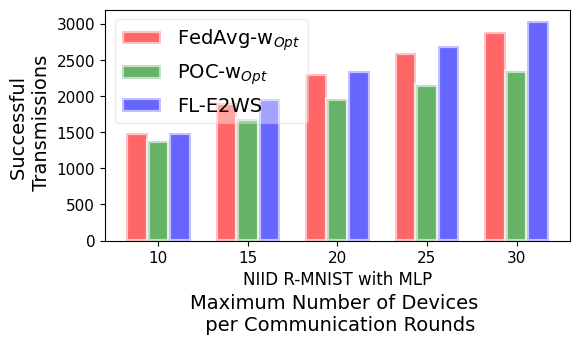}
\label{mnist-num-disp}
}
\hfil
\subfloat[]{\includegraphics[width=0.45\textwidth]{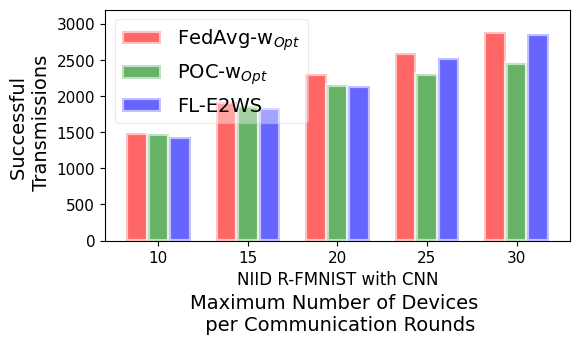}
\label{fmnist-num-disp}
}
\caption{Number of successful transmissions with $10 \leq n_f \leq 30$. In all variations of $n_f$, FL-E2WS seeks to minimize energy consumption, and yet maintains the average number of transmissions compared to other algorithms that operate with bandwidth allocation and fixed power.}
\label{fig:num-disp}
\end{figure}

Figures \ref{mnist-num-disp} and \ref{fmnist-num-disp} show the number of successful transmissions after $200$ rounds of communication, varying the value of $n_f$, which defines the maximum number of devices in each round of communication. As the value of $n_f$ increases, the number of successful transmissions also increases. In all variations of $n_f$, FL-E2WS seeks to minimize energy consumption and yet maintains the average number of transmissions compared to other algorithms that operate with fixed bandwidth and power allocation.

The difference in the number of successful transmissions is characterized by the configuration of the strategies of the different algorithms. In the context of communication resource allocation, it is noteworthy that FL-E2WS seeks energy efficiency by minimizing the transmission power of the devices, causing the probability of successful transmission of the models to be slightly lower compared to other strategies that operate stably with fixed power. Furthermore, it is noteworthy that FL-E2WS is capable of achieving a significantly greater number of transmissions when energy efficiency is not a requirement to be considered.

Regarding device selection strategies, FL-E2WS and FedAvg-w$_{Opt}$ are similar in their contribution to the metric related to the number of successful transmissions. While FedAvg-w$_{Opt}$ randomly determines the selection of devices, FL-E2WS also randomly selects a partial number of devices and then determines the final $n_f$ quantity of devices for each round of communication, directing the strategy towards local training that covers more data to improve the representativeness of locally trained models. In this case, the impact of different strategies must be observed on the accuracy of the global model.

The average number of successful transmissions of \mbox{POC-w$_{Opt}$} is lower compared to the other algorithms. In a wireless networking context, POC-w$_{Opt}$ may underperform due to its device selection strategy based on average accumulated loss, which selects devices with the highest values of the local loss function. The initialization of \mbox{POC-w$_{Opt}$} assumes that the value of the loss function is high for devices that have not yet been successful in transmitting local models to the aggregator server. This can lead to a scenario in which POC-w$_{Opt}$ insists on selecting devices that are unable to transmit their models successfully due to wireless channel conditions, leading to increased transmission errors.

\begin{figure*}[!t]
\centering
\subfloat[]{\includegraphics[width=0.42\textwidth]{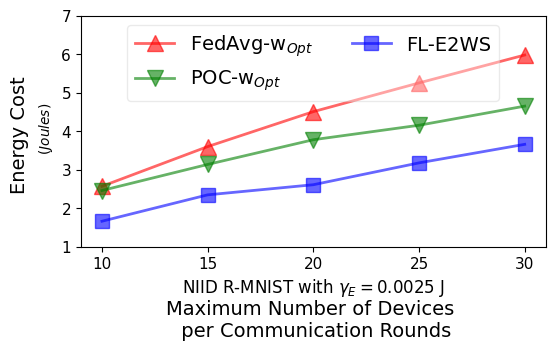}
\label{custo-energetico-mnist}}
\hfil
\subfloat[]{\includegraphics[width=0.42\textwidth]{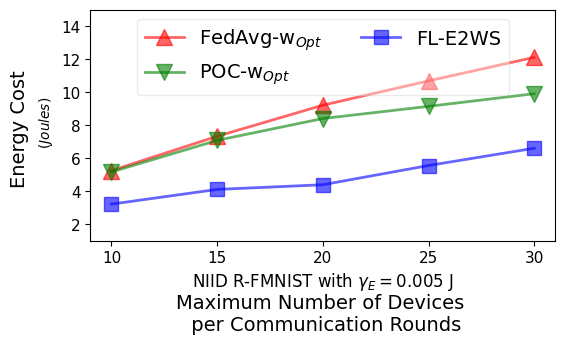}
\label{custo-energetico-fmnist}}
\caption{Energy cost with $10 \leq n_f \leq 30$. The results show that FL-E2WS is in accordance with the formulation of the objective function of the Equation (\ref{eq:escalonamento-recursos}) which aims to minimize the energy consumption of the devices.}
\label{fig_mnist_acuracia}
\end{figure*}

Figures \ref{custo-energetico-mnist} and \ref{custo-energetico-fmnist} show the total energy cost after $200$ rounds of communication, varying the value of $n_f$. It is observed that the energy cost increases almost linearly as the maximum number of selected devices increases and the proportion of successful transmissions of each algorithm. Furthermore, it is possible to note that FL-E2WS is efficient in minimizing the use of energy resources used in training and transmitting local models to the aggregator server, setting a lower bound in relation to the other algorithms. This is due to the fact that energy consumption is directly related to the transmission power of the devices and the transmission time of local models. In this case, the results are in accordance with the formulation of the objective function in Equation (\ref{eq:escalonamento-recursos}) of FL-E2WS, where the aim is to minimize power and transmission time while maximizing the number of devices selected in each round of communication.

\begin{figure}[!t]
\centering
\subfloat[]{\includegraphics[width=0.45\textwidth]{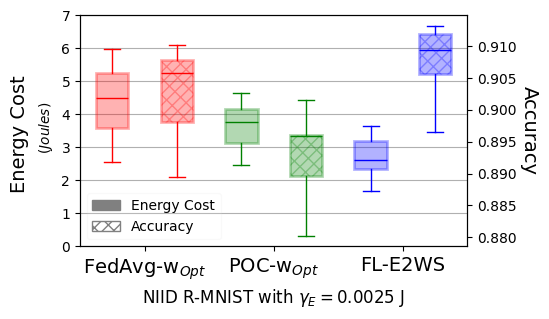}
\label{mnist-energia-bp}
}
\hfil
\subfloat[]{\includegraphics[width=0.45\textwidth]{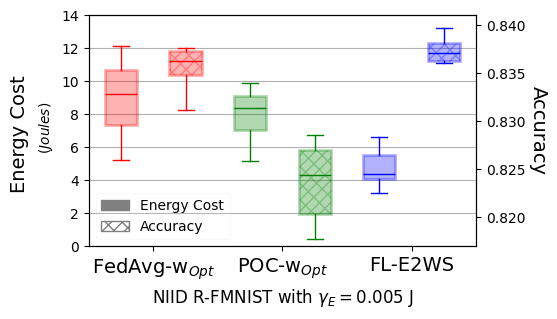}
\label{fmnist-energia-bp}
}
\caption{Distribution of energy cost and accuracy with $10 \leq n_f \leq 30$. The results of the energy cost distribution and accuracies show that FL-E2WS is energy efficient while ensuring the convergence of the global model.}
\label{fig:num-disp}
\end{figure}

Figures \ref{mnist-energia-bp} and \ref{fmnist-energia-bp} present diagrams with the main statistics of the distribution of the total energy cost and the accuracy of the algorithms after $200$ rounds of communication, varying the value of $n_f$. The solidly filled boxes show the statistical summary of the energy cost, and the crossed-out boxes show the statistical summary of the accuracy.

The median of the energy cost distribution of FL-E2WS is more towards the center of the box, indicating a symmetrical distribution. In the case of POC-w$_{Opt}$, the median of the distribution is slightly above the center of the box, indicating that values tend to be higher. For FL-E2WS, the median of the energy cost distribution is closer to the bottom of the box, indicating that the values tend to be lower. On the other hand, considering the scale of the accuracy distribution, the performance of the global model of the algorithms is similar. However, FL-E2WS performs slightly better when looking at the median, minimum, and maximum accuracies. Thus, the results of the energy cost distribution and accuracies show that FL-E2WS is energy efficient while ensuring the convergence of the global model.

\section{Conclusion and Future Work}
\label{conclusions}

This work investigated the problem of device selection and communication resource allocation in wireless IoT networks for FL tasks. Therefore, FL-E2WS was formulated as an optimization problem that considers factors inherent to the wireless environment along with factors that influence the model learning process. The dynamic allocation of FL-E2WS communication resources allowed the minimization of the devices energy consumption while maintaining a similar number of transmissions compared to other algorithms that operate with fixed bandwidth and transmission power. At the same time, FL-E2WS achieved slightly better global accuracy by targeting devices for local training that covered a broader set of data. Furthermore, it should be noted that the source code of FL-E2WS is available to allow the reproduction and validation of the results. For future work, we intend to investigate other techniques for optimizing different FL architectures in wireless IoT networks, aiming to improve the performance of the global model, minimize energy consumption, and consider device mobility.

\begin{comment}
%\newpage

\section{Biography Section}
If you have an EPS/PDF photo (graphicx package needed), extra braces are
 needed around the contents of the optional argument to biography to prevent
 the LaTeX parser from getting confused when it sees the complicated
 $\backslash${\tt{includegraphics}} command within an optional argument. (You can create
 your own custom macro containing the $\backslash${\tt{includegraphics}} command to make things
 simpler here.)
 
\vspace{11pt}

\bf{If you include a photo:}\vspace{-33pt}
\begin{IEEEbiography}[{\includegraphics[width=1in,height=1.25in,clip,keepaspectratio]{fig1}}]{Michael Shell}
Use $\backslash${\tt{begin\{IEEEbiography\}}} and then for the 1st argument use $\backslash${\tt{includegraphics}} to declare and link the author photo.
Use the author name as the 3rd argument followed by the biography text.
\end{IEEEbiography}

\vspace{11pt}

\bf{If you will not include a photo:}\vspace{-33pt}
\begin{IEEEbiographynophoto}{John Doe}
Use $\backslash${\tt{begin\{IEEEbiographynophoto\}}} and the author name as the argument followed by the biography text.
\end{IEEEbiographynophoto}
\end{comment}

\vfill

\end{document}